\documentclass[aps,prl,twocolumn,amssymb,showpacs]{revtex4}
\newcommand{\fma}{f_{\rm max}}
\newcommand{\ema}{\epsilon''_{\rm max}}
\newcommand{\fmo}{f_0}
\newcommand{\emo}{\epsilon''_0(T)}
\usepackage{graphicx}
\begin{document}
\title{Minimal model for beta relaxation in viscous liquids}
\author{Jeppe C. Dyre and Niels Boye Olsen}
\affiliation{Department of Mathematics and Physics (IMFUFA), 
Roskilde University, Postbox 260, DK-4000 Roskilde, DENMARK}
\date{\today}

\begin{abstract}
Contrasts between beta relaxation in equilibrium viscous liquids and glasses are 
rationalized in terms of a double-well potential model with 
structure-dependent asymmetry, assuming structure is described by a single order 
parameter. The model is tested for tripropylene glycol where it accounts for the 
hysteresis of the dielectric beta loss peak frequency and magnitude during cooling 
and reheating through the glass transition.
\end{abstract}
\maketitle

Viscous liquids approaching the calorimetric glass transition have extremely long 
relaxation times \cite{viscliq}.  The main relaxation is termed the alpha relaxation. 
There is usually an additional minor ``beta'' process at higher frequencies. Dielectric 
relaxation is a standard method for probing liquid dynamics \cite{diel}. The study of 
dielectric beta relaxation in simple viscous liquids was pioneered by Johari and 
Goldstein more than 30 years ago \cite{joh70,joh85}, but the origin of beta 
relaxation is still disputed \cite{kah97,wag98,joh02a}. It is unknown whether every 
molecule contributes to the relaxation \cite{all} or only those within ``islands of 
mobility'' \cite{gol69,joh76,kop00}. Similarly, it is not known whether small angle 
jumps \cite{all,kau90,vog00} or large angle jumps \cite{arb96} are responsible for 
the beta process.

Improvements of experimental techniques have recently lead to several new 
findings. The suggestion \cite{ols98,leo99,wag99} that the excess wing of the alpha 
relaxation usually found at high frequencies is due to an underlying low-frequency 
beta process was confirmed by long time annealing experiments by Lunkenheimer 
and coworkers \cite{lunken} (an alternative view is that the wing is a non-beta type 
relaxation process \cite{paluch}). This lead to a simple picture of the alpha process: 
Once the effect of interfering beta relaxation is eliminated, alpha relaxation obeys 
time-temperature superposition with a high frequency loss $\propto\omega^{-1/2}$ 
\cite{ols01}. Moreover, it now appears likely that all liquids have one or more beta 
relaxations \cite{lunken,han97,kud97,joh02b}: Liquids like propylene carbonate, 
glycerol, salol, and toluene are now known to possess beta relaxation, while 
o-terphenyl, previously thought to have a beta process only in the glassy state, has 
one in the equilibrium liquid phase as well. Finally, it has been shown that beta 
relaxation in the equilibrium liquid does not behave as expected by extrapolation 
from the glassy phase: In some cases the beta loss peak frequency is temperature 
independent in the liquid phase (e.g., sorbitol \cite{ols98}), in other cases it is very 
weakly temperature dependent. On the other hand, the beta relaxation strength 
always increases strongly with temperature in the liquid phase \cite{ols00,note1}.

The contrasts between beta relaxation in liquid and glass are clear from Fig. 1(a) 
which shows beta loss peak frequency and maximum loss for tripropylene glycol 
cooled through the glass transition and subsequently reheated \cite{note2}.
\begin{figure}
\includegraphics[scale=0.45]{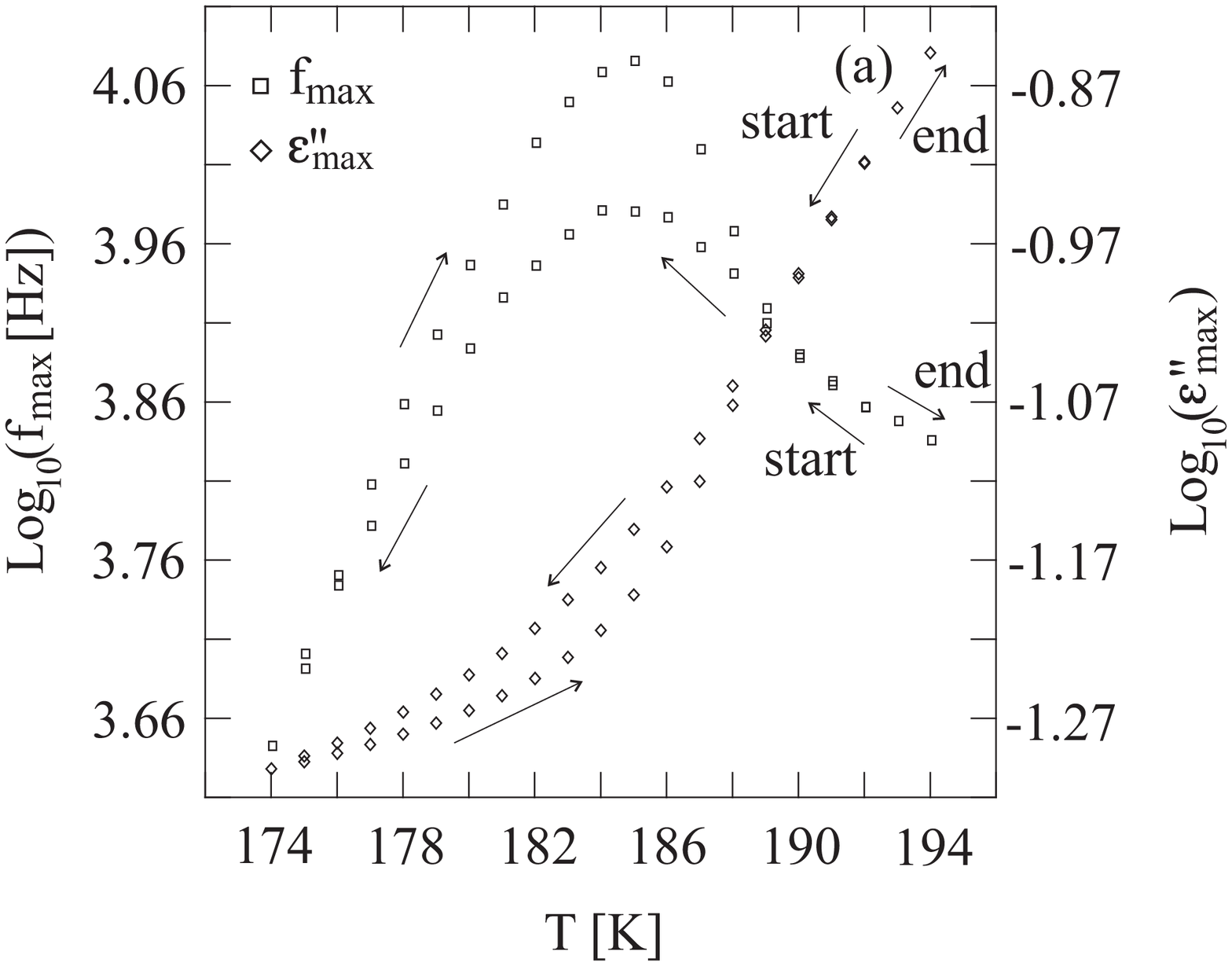}
\includegraphics[scale=0.45]{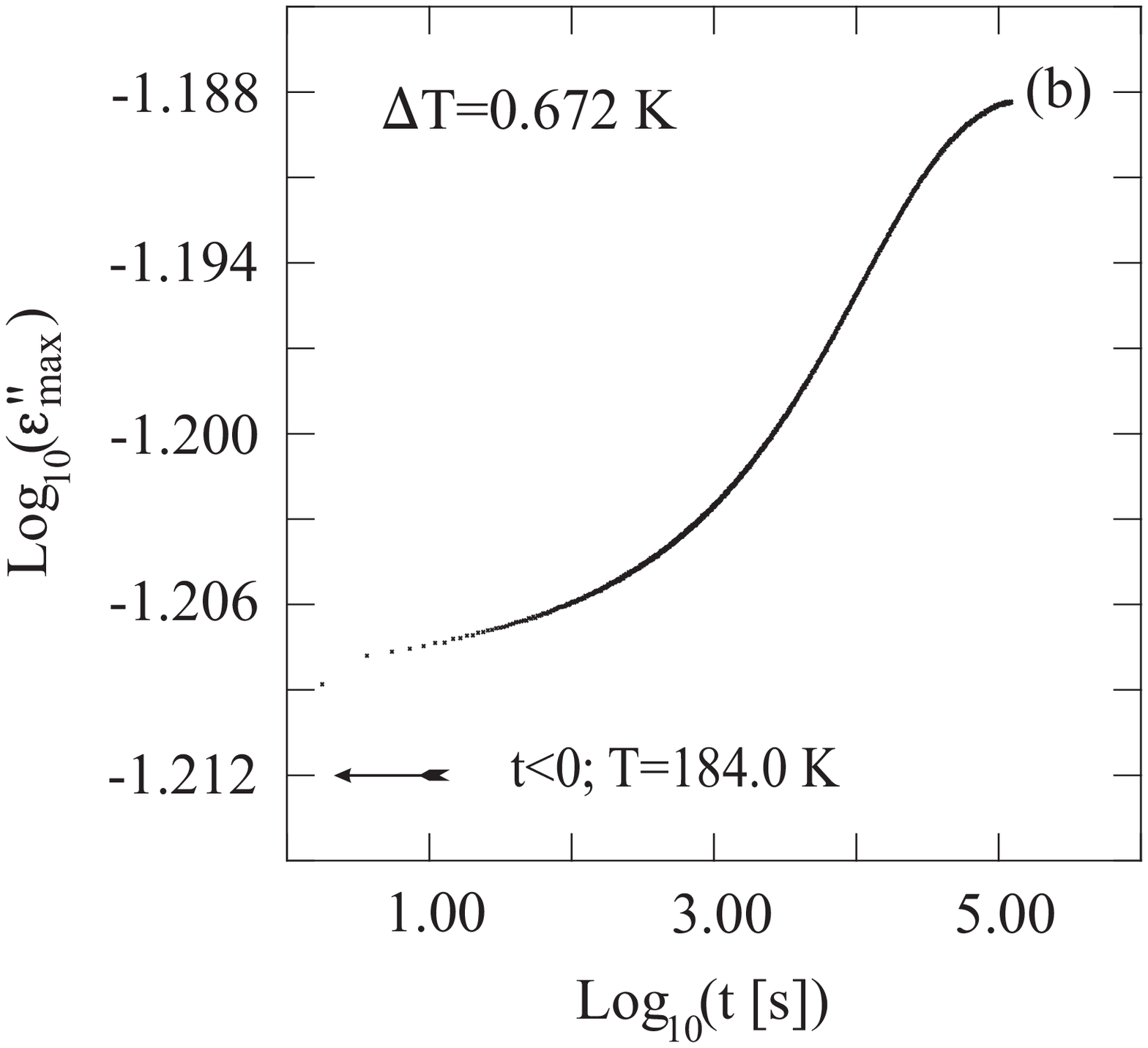}
\caption{Data for tripropylene glycol. (a) Observation of beta loss peak frequency 
($\square$) and loss peak maximum ($\Diamond$) for a continuous passage through 
the glass transition starting from the equilibrium viscous liquid at 192 K, cooling to 
174 K, and reheating to 194 K. After each temperature step the systems is kept at 
constant temperature for 50 min whereafter a spectrum is measured (sampling time: 6 
min). In the glassy phase one observes the well-known strongly 
temperature-dependent loss peak frequency and weakly 
temperature-dependent maximum loss; in the liquid phase the situation is the 
opposite (as seen also in other liquids, e.g., sorbitol \cite{ols98,ols00}). There is even 
a reversal so that the loss peak frequency decreases with increasing temperature. (b) 
Beta loss peak maximum monitored after a temperature increase of 0.672 K starting 
from equilibrium at 184.0 K. After 6 s temperature is stable within 1 mK of the final 
temperature. 
\label{fig1}}
\end{figure}
In the glassy phase (at low temperatures) the loss peak frequency is strongly 
temperature dependent while the maximum loss varies little. On the other hand, 
the loss is strongly temperature dependent in the equilibrium liquid phase. Here we 
even see the loss peak frequency decreasing upon heating. How is one to understand 
these findings? A clue is provided by Fig. 1(b) which shows the maximum loss as a 
function of time after an  ``instantaneous'' temperature step, i.e., instantaneous on the 
time scale of structural (alpha) relaxation. This experiment utilizes a special-purpose 
setup with a cell consisting of two aluminum discs separated by three capton spacers 
(layer distance $\sim\rm 20\mu$, empty capacitance 30 pF). One disc, where 
temperature is measured via an NTC resistor, is placed on a Peltier element. Less 
than six seconds  after a 0.672 K temperature jump is initiated, temperature is stable 
within 1 mK. In this setup we measure at 10 kHz which is the loss peak frequency 
(changes of loss peak frequency lead only to second order corrections of $\ema$). 
The sampling time is 2 s. Figure 1(b) shows a very fast change of the maximum loss, 
followed by relaxation toward the equilibrium value taking place on the structural 
(alpha) relaxation time scale. The existence of an instantaneous increase of the loss 
clearly indicates a pronounced asymmetry of the relaxing entity. Inspired by this fact 
we adopt the standard asymmetric double-well potential model (Fig. 2) with 
transitions between the two free energy minima described by rate theory 
\cite{diel,glref,pol72,gil81,buc01}.
\begin{figure}
\includegraphics[scale=0.45]{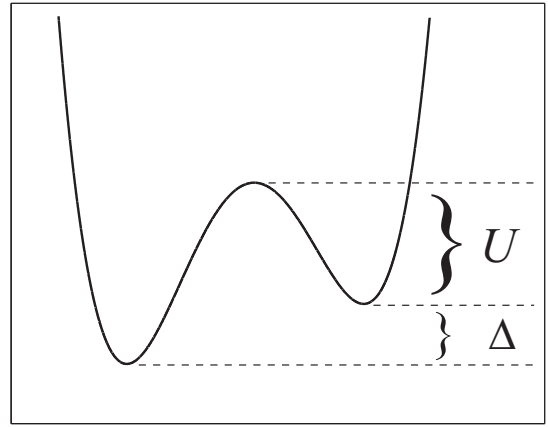}
\caption{Asymmetric double well potential as a simple model for beta relaxation. 
The two free energy differences $U$ and $\Delta$ vary as structure changes with 
temperature in the equilibrium liquid phase, but freeze at the glass transition. 
Working from this picture the simplest possible assumptions are that 1) Structure is 
parameterized by just one order parameter, and 2) first order Taylor expansions apply 
in the relatively narrow temperature range studied. 
\label{fig2}} 
\end{figure}
In terms of the small barrier $U$ and the asymmetry $\Delta$, loss peak frequency 
$\fma$ and maximum loss $\ema$ are given \cite{gil81} by

\begin{equation}\label{1}
\left\{
\begin{array}{ccl}
\fma & = & \fmo \ \displaystyle
\exp\left(-\frac{2U+\Delta}{2k_BT}\right)\ 
\cosh\left(\frac{\Delta}{2k_BT}\right)\\
\ema & = & \emo \ \displaystyle
\cosh^{-2}\left(\frac{\Delta}{2k_BT}\right)
\end{array}
\right.\ \ .
\end{equation}
The prefactor $\fmo$ is assumed to be structure and temperature independent while 
$\emo\propto 1/T$ \cite{gil81} is assumed to be structure independent: 
$\emo=T_0/T$. The free energy differences $U$ and $\Delta$ are expected to 
change with changing structure, but freeze at the glass transition. In terms of the 
fictive temperature $T_f$ our model is based on 

\begin{equation}\label{2}
\left\{
\begin{array}{ccccc}
U      & = &  U_0  &  + & a\, k_B T_f\\
\Delta & = &  \Delta_0 & - & b\, k_B T_f
\end{array}
\right.\ \ .
\end{equation}
Equation (\ref{2}) follows from minimal assumptions: Suppose structure is 
parameterized by just one variable, $s$. Only a rather narrow range of temperatures 
is involved in studies of beta relaxation in the liquid phase and around the glass 
transition. Consequently, structure varies only little and, e.g., $U(s)$ may be 
expanded to first order: $U(s)=c_0+c_1(s-s_0)$. For the equilibrium liquid, $s=s(T)$ 
which may also be expanded to first order. By redefining $s$ via a linear 
transformation we obtain $s=T$ at equilibrium while $U(s)$ is still linear in $s$. A 
single variable describing structure, which at equilibrium is equal to temperature, is 
-- consistent with Tool's 1946 definition \cite{too46} -- to be identified with the 
fictive temperature: $s=T_f$. Thus one is lead to Eq. (\ref{2}) where signs are 
chosen simply to ensure $a,b>0$ in fit to data.

The model has 6 parameters: $\fmo$, $T_0$, $U_0$, $\Delta_0$, $a$, and $b$. 
These were determined by measuring the instantaneous changes of $\fma$ and 
$\ema$ upon a temperature step, as well as their thermal equilibrium changes. Figure 
1(b) allows determination of the instantaneous change of the loss. It is not possible to 
determine the  instantaneous change of $\fma$. Instead we extrapolated 
measurements obtained by the standard cell \cite{note2} as follows (Fig. 3): Beta 
loss is monitored by first annealing at 183.0 K, subsequently changing temperature to 
181.0 K. The latter data show a linear relation between $\log\fma$ 
and $\log\ema$ which, knowing the instantaneous change per Kelvin of $\ema$ from 
Fig. 1(b), is extrapolated to short times. 

In the data analysis it is convenient to eliminate $\cosh$ by introducing the variable

\begin{equation}\label{3}
Y\ \equiv\ \left(\frac{\fma}{\fmo}\right)^2\ \left(\frac{\ema}{\emo}\right)\,.
\end{equation}
On the fast time scale $U$ and $\Delta$ are frozen so  Eq. (\ref{1}) implies 

\begin{equation}\label{4}
\left\{\begin{array}{ccc}\displaystyle
\left.\frac{d\ln Y}{d\ln T}\right|_{\rm inst} & = &
\displaystyle\frac{2U+\Delta}{k_BT}\\ \displaystyle
\left.\frac{d\ln\ema}{d\ln T}\right|_{\rm inst}  & = &
\displaystyle
\frac{\Delta}{k_BT}\ \sqrt{1-\frac{\ema}{\emo}}\ -\ 1
\end{array}
\right. \ \ .
\end{equation}
\begin{figure}
\includegraphics[scale=0.45]{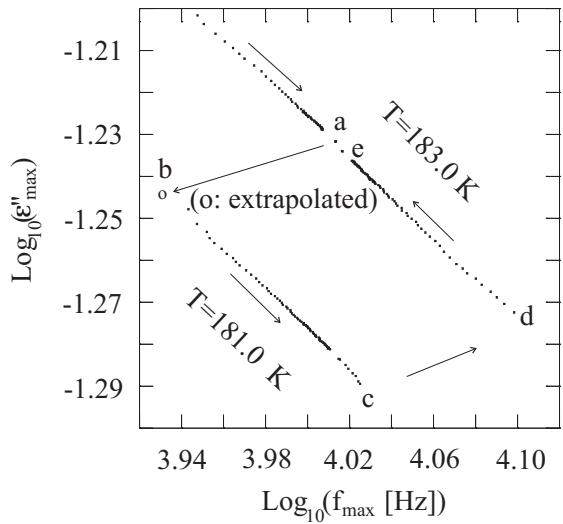}
\caption{Reversible temperature jump experiment for tripropylene glycol monitoring 
beta loss peak frequency and loss peak maximum. Starting at 185.0 K (not shown) 
temperature is first lowered to 183.0 K and kept there for 60 h; the last point, a, was 
obtained after further 24 h. Then temperature is changed to 181.0 K where it is kept 
constant for 140 h (initially 20 min between measurements, later 12 hours). The point 
b is found by extrapolating to zero time after the quench utilizing the data of Fig. 
1(b). This series ends at point c, thereafter temperature is changed back to 183.0 K 
where it is kept for 140 h (all points except the final point e refer to times up to 70 h; 
e was measured after further 70 h). Equilibrium at 183.0 K is somewhere between 
points a and e.} 
\end{figure}
Having determined the instantaneous changes of $\ema$ and $\fma$, Eqs. (\ref{1}) 
and (\ref{4}) provide 4 equations for the 6 model parameters. The last two equations 
come from the temperature dependence of loss and loss peak frequency at thermal 
equilibrium where $U(T)$ and $\Delta(T)$ are given by Eq. (\ref{2}) with $T_f=T$, 
leading to 

\begin{equation}\label{5}
\left\{\begin{array}{ccc}\displaystyle
\left.\frac{d\ln Y}{d\ln T}\right|_{\rm eq} & = &
\displaystyle
\frac{2U_0+\Delta_0}{k_BT}\\ \displaystyle
\left.\frac{d\ln\ema}{d\ln T}\right|_{\rm eq}  & = &
\displaystyle
\frac{\Delta_0}{k_BT}\ \sqrt{1-\frac{\ema}{\emo}}\ -\ 1
\end{array}
\right. \ \ .
\end{equation}
Using Eq. (\ref{4}) for the data of Figs. 1(b) and 3, and using Eqs. (\ref{1}) and 
(\ref{5}) for equilibrium state measurements at 183 and 185 K, the 6 parameters 
determined for tripropylene glycol \cite{note3} are: 
$\fmo=6.2\cdot 10^{11}$ Hz, 
$T_0=130$ K, 
$U_0/k_B=-1107$ K, 
$\Delta_0/k_B=3039$ K, 
$a=23.28$, 
$b=12.77$. 
$U_0<0$ reflects the fact that the beta loss peak frequency in the liquid phase 
decreases as temperature increases [Fig. 1(a)]. Physically, this anomalous behavior is 
caused by the barrier $U$ increasing more than $T$ upon heating.

Once all parameters are fixed the model predicts how $\fma$ and $\ema$ correlate 
for the continuous passage through the glass transition of Fig. 1(a). To analyze these 
data within the model we first note that if $y(x)=1/\cosh^2(x)$ then $x=\phi(y)$ 
where 
$\phi(y)=\ln(1/\sqrt{y}+\sqrt{1/y-1})$, so Eq. (\ref{1}) may be inverted: 
$\Delta/2k_BT=\phi(\ema/\emo)\equiv\Phi $. Since 
$\ln Y=-(2U+\Delta)/k_BT$ both $\Phi$ and $\ln Y$ involve fictive temperature, and consequently both exhibit hysteresis at the glass transition. However, fictive 
temperature is eliminated by considering the following variable

\begin{eqnarray}\label{6}
Z(\fma,\ema)\ & \equiv\ & \Big(2a-b\Big)\Phi-\frac{b}{2}\ln\ Y\nonumber\\ 
& = & \ \frac{a\ \Delta_0\ +\ b\ U_0}{k_BT}\ \ .
\end{eqnarray}
The model is tested in Fig. 4. The glass transition is not visible and most hysteresis is eliminated 
(better elimination was obtained in Ref. \cite{ols00} but without theoretical basis and 
with one free parameter \cite{note4}). The line shown is the prediction of Eq. 
(\ref{6}).
\begin{figure}
\includegraphics[scale=0.45]{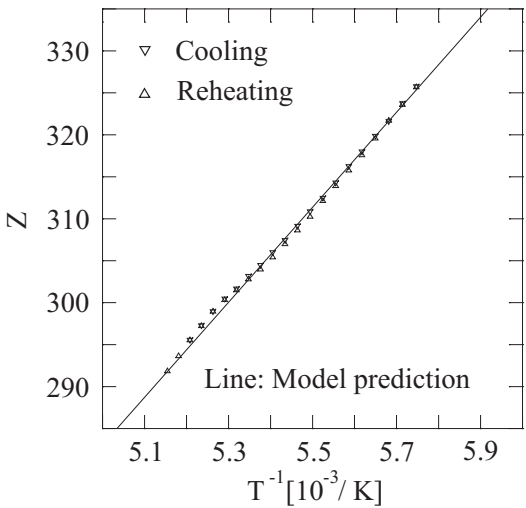}
\caption{Data for tripropylene glycol from Fig. 1(a) replotted to test the model. The 
variable $Z$ defined in Eq. (\ref{6}) is calculated using parameters obtained by independent experiments.
\label{fig4}}
\end{figure}

In conclusion, a minimal model for beta relaxation in viscous liquids has been 
proposed. The model is built on the four simplest possible assumptions: 1) Beta 
relaxation involves only two levels, 2) structure is determined by just one 
order-parameter, 3) first order Taylor expansions apply, 4) the two characteristic free 
energies $U$ and $\Delta$ freeze at the glass transition. The model is clearly 
oversimplified. For instance, it predicts a Debye response which is not observed, and 
$U$ and $\Delta$ would be expected to vary slightly with temperature in the glass. 
Nevertheless, the model is able to rationalize the contrasts between beta relaxation in 
liquids and in glasses. One final puzzling observation should be mentioned: The 
asymmetry $\Delta$ extrapolates to zero at a temperature which is close to the 
temperature where alpha and beta relaxations merge. We have seen the same 
phenomenon in sorbitol, a pyridine-toluene solution, 
polypropylene-glycol-425, and in 
4,7,10-trioxydecane-1,13-diamine \cite{ols98,ols00}, and have found no exceptions. 
This finding indicates that the merging temperature is fundamental, a symmetry is 
somehow broken below this temperature.

\acknowledgments
 This work was supported by the Danish Natural Science Research Council.


\begin{thebibliography}{99}
\bibitem{note2} The experimental set-up is identical to that used in Ref. \cite{ols00},  
except for an improved non-commercial 14-bit cosine-wave generator at 16 fixed 
frequencies per decade used for frequencies below 100 Hz. To eliminate the 
influence from the high frequency alpha tail the data were subjected to the following 
procedure:  First we obtained the equilibrium shape of the beta peak by annealing for 
100 h at 184 K. Then, assuming invariance of the shape of the beta peak, the alpha 
tail influence was eliminated by subtracting a term $\propto\omega^{-1/2}$ 
\cite{ols01} with constant of proportionality adjusted to obtain the invariant shape of 
the beta peak. This procedure is unique. The subtraction procedure mainly affects 
$\fma$ and only influences measurements at the highest temperatures. The observed 
hysteresis effects are much larger than the corrections from subtracting the alpha tail 
\cite{ols00}.
\end{thebibliography}
\end{document}